\documentclass[aps,pra,twocolumn]{revtex4}
\usepackage[dvips]{epsfig}
\usepackage[usenames]{color}
\usepackage{pstricks}
\usepackage{amsmath}
\usepackage{amssymb}
\usepackage{subfigure}

\begin{document}

\title{Strong Dependence of Ultracold Chemical Rates on Electric Dipole Moments}

\author{Goulven Qu{\'e}m{\'e}ner and John L. Bohn}
\affiliation{JILA, University of Colorado,
Boulder, C0 80309-0440, USA}

\date{\today}

\begin{abstract}
We use the quantum threshold laws combined with a classical capture model
to provide an analytical estimate
of the chemical quenching cross sections and rate coefficients of two colliding particles
at ultralow temperatures.
We apply this quantum threshold model (QT model) to indistinguishable fermionic polar molecules in an electric field.
At ultracold temperatures and in weak electric fields, the cross sections and rate coefficients depend
only weakly
on the electric dipole moment $d$ induced by the electric field.
In stronger electric fields, 
the quenching processes scale as $d^{4(L+\frac{1}{2})}$
where $L>0$ is the orbital angular momentum quantum number between the two colliding particles.
For $p-$wave collisions ($L=1$) of indistinguishable fermionic polar molecules at ultracold temperatures,
the quenching rate thus scales as $d^6$.
We also 
apply this model to pure two dimensional collisions
and find that
chemical rates vanish as $d^{-4}$
for ultracold indistinguishable fermions.
This model provides a quick and intuitive way to estimate
chemical rate coefficients of reactions occuring with high probability.
\end{abstract}


\maketitle

\font\smallfont=cmr7

\section{Introduction}

Ultracold samples of bi-alkali polar molecules have been created
very recently in their ground electronic
$^1\Sigma$, vibrational $v=0$, and 
rotational $N=0$ states~\cite{Sage05,Ni08,Deiglmayr08}. This is a
promising step before achieving Bose-Einstein condensates or
degenerate Fermi gases of polar molecules, provided that
further evaporative cooling is efficient. For this purpose, elastic
collision rates  must be much faster than inelastic quenching
rates.
This issue is somewhat problematic for
the bi-alkali molecules recently created, since they are subject to
quenching via chemical reactions.  If a reaction should occur, 
the products are no longer trapped.

 For alkali dimers that possess
electric dipole moments, elastic scattering appears to be quite
favorable, since elastic scattering rates are expected to scale with the 
fourth power of the dipole moment~\cite{Hensler03,Bohn09}.
Inelastic collisions of polar species can originate from two
distinct sources.  The long-range dipole-dipole
interaction itself is anisotropic and can cause dipole orientations
to be lost.  This kind of loss generally leads to high inelastic rates,
and is regarded as the reason why electrostatic trapping of polar
molecules is likely not feasible~\cite{Avdeenkov02}.  Moreover,
these collisions also scale as the fourth power of dipole moment in
the ultracold limit~\cite{Hensler03}, meaning that the ratio of 
elastic to inelastic rates does not in general improve at higher 
electric fields.  This sort of loss can be prevented by trapping the
molecules in optical dipole traps.

More serious is the possibility that collisions are quenched by 
chemical reactions.  Chemical reaction rates are known to be
potentially quite high even at ultracold 
temperatures~\cite{Bala01,Soldan02,Quemener04,Quemener05,Cvitas05a,Cvitas05b,Lara06,Quemener07,Hutson07,QuemenerCHAPTER,Quemener09}.
Indeed, for collision energies
above the Bethe--Wigner threshold regime, it appears that many quenching rates,
chemical or otherwise, of barrierless systems are well described by applying Langevin's
classical model~\cite{Langevin05}.  In this model the molecules
must surmount a centrifugal barrier to pass close enough to react,
but are assumed to react with unit probability when they do so.
This model has adequately described several cold molecule 
quantum dynamics calculations~\cite{Quemener05,Cvitas05a,Lara06,QuemenerCHAPTER,Quemener09}.

Within the Bethe--Wigner limit, scattering can be described by an elegant
Quantum Defect Theory (QDT) approach~\cite{Julienne89,Burke98,Mies00,Julienne09}.
This approach makes explicit
the dominant role of long-range forces in controlling how likely the
molecules are to approach close to one another.  Consequently, quenching rate
constants can often be written in an analytic form that contains a small
number of parameters that characterize short-range physics such as 
chemical reaction probability.  For processes in which the
quenching probability is close to unity, the QDT theory provides
remarkably accurate quenching rates~\cite{Orzel99,Hudson08}.
For dipoles, however, the full QDT theory remains to be formulated.

In this article we combine  two powerful ideas -- suppression
of collisions due to long-range physics, and high-probability
quenching inelastic collisions for those that are not suppressed -- to derive
simple estimates for inelastic/reactive scattering rates for
ultracold fermionic dipoles.  
The theory arrives at
remarkably simple expressions of collision rates, without
the need for the full machinery of close-coupling calculations.
Strikingly, the model shows that quenching collisions scale as the
{\it sixth} power of the dipole moment for ultracold $p-$wave collisions.  
On the one hand, this implies
a tremendous degree of control over chemical reactions by simply
varying an electric field, complementing alternative proposals
for electric field control of
molecule-molecule~\cite{Hudson06} or atom-molecule~\cite{Tscherbul08} 
chemistry.
On the other hand, it also implies that
evaporative cooling of polar molecules may become more difficult
as the field is increased.
In section II, we formulate the theoretical model for three dimensional
collisions. In section III, we apply this model to pure two dimensional 
collisions and conclude in section IV.
In the following, quantities are expressed in S.I. units, unless explicitly stated otherwise.
Atomic units (a.u.) are obtained by setting $\hbar = 4 \pi \varepsilon_0 = 1$.

\section{Collisions in three dimensions}

\subsection{Cross sections and collision rates}

In quantum mechanics, the quenching cross section of a pair of colliding molecules (or any particles) 
of reduced mass $\mu$
for a given collision energy $E_c$ and a partial wave $L,M_L$ is given by
\begin{eqnarray}
\sigma^{qu}_{L,M_L} &=&  \frac{\hbar^2  \pi}{2 \mu E_c} \ |T^{qu}_{L,M_L}|^2  \times \Delta  
\label{cross}
\end{eqnarray}
where $T^{qu}$ is the transition matrix element of the quenching process,
$|T^{qu}_{L,M_L}|^2$ represents the quenching probability,
and the factor $\Delta$ 
represents symmetrization requirements for identical particles~\cite{Burke99}.
If the two colliding molecules are in different internal quantum states (distinguishable molecules), $\Delta=1$
and if the two colliding molecules are in the same internal quantum state (indistinguishable molecules), $\Delta=2$.
The total quenching cross section of a pair of molecules
is $\sigma^{qu} = \sum_{L,M_L} \sigma^{qu}_{L,M_L} $. 
The quenching rate coefficient of a pair of molecules for a given temperature $T$ (collisional event rate)
is given by
%
\begin{eqnarray}
K^{qu}_{L,M_L} =  < \sigma^{qu}_{L,M_L} \times  v > 
= \int_0^\infty \sigma^{qu}_{L,M_L} \,  v \, f(v) \, dv
\label{rate}
\end{eqnarray}
%
where 
\begin{eqnarray}
f(v)  = 4 \pi \, \left( \frac{\mu}{2 \pi k_B T} \right)^{3/2} \, v^2 \, \exp{[-(\mu v^2)/(2 \, k_B \, T)]} 
\end{eqnarray}
%
is the Maxwell--Boltzmann distribution for the relative velocities for a given temperature
and $k_B$ is the Maxwell--Boltzmann constant.
The total quenching rate coefficient of a pair of molecules
is $K^{qu} = \sum_{L,M_L} K^{qu}_{L,M_L} $.
To avoid confusion, we will also write the corresponding
rate equation for collisions between distinguishable and
indistinguishable molecules. 
First, we consider collisions between two distinguishable molecules in quantum states $a$ and $b$ 
($\Delta=1$ in Eq.~\eqref{cross}).
During a time $dt = \tau$, where $\tau$ is the time 
of a quenching collisional event, 
the number of molecules $N_a$ lost in each collision is one and 
the number of molecules $N_b$ lost in each collision is one. 
Then $dN_a/dt=-1/\tau$ and $dN_b/dt=-1/\tau$.
The volume per colliding pairs of molecules is $V/(N_a \, N_b)$, 
where $V$ stands for the volume of the gas.
During the time $\tau$, the quenching collisional event
is associated with a volume $<\sigma^{qu} \times v> \times \tau = K^{qu} \times \tau $.
By definition of $\tau$, this volume should be equal to the one 
occupied by just one colliding pair of molecules.
Then we get $K^{qu} \times \tau  = V/(N_a \, N_b)$.
The rate equation for the number of molecule $N_a$ or $N_b$ 
is then given by
\begin{eqnarray}
\frac{dN_{a,b}}{dt} = - K^{qu} \times \frac{N_a \, N_b}{V} .
\end{eqnarray}
If $n_a = N_a / V$ and $n_b = N_b / V$ are the densities of molecule $a$ and $b$ in the gas, then
\begin{eqnarray}
\frac{dn_{a,b}}{dt} = - K^{qu} \times n_a \, n_b.
\end{eqnarray}
We consider now the case of
collisions between two indistinguishable molecules   
($\Delta=2$ in Eq.~\eqref{cross}).
During the time $dt = \tau$
the number of molecules $N$ lost in each collision is two.
Then we get $dN/dt=-2/\tau$. 
The volume per colliding pairs of molecules is $V/(N(N-1)/2)$
where we have taken into account the indistinguishability of the molecules.
For the same reason explained above, 
the volume associated with the collisional event
during the time $\tau$
should be equal to the volume 
occupied by just one colliding pair of molecules.
And then $K^{qu} \times \tau  = V/(N(N-1)/2)$.
The rate equation for the number of molecule $N$ 
is then given by
\begin{eqnarray}
\frac{dN}{dt} = - 2 \, K^{qu} \times \frac{N(N-1)/2}{V} .
\end{eqnarray}
If $n = N / V$ and $N(N-1) \approx N^2$, then
\begin{eqnarray}
\frac{dn}{dt} = - K^{qu} \times n^2.
\end{eqnarray}

\subsection{Quantum threshold model}

\begin{figure} [t]
\begin{center}
\includegraphics*[width=8cm,keepaspectratio=true,angle=0]{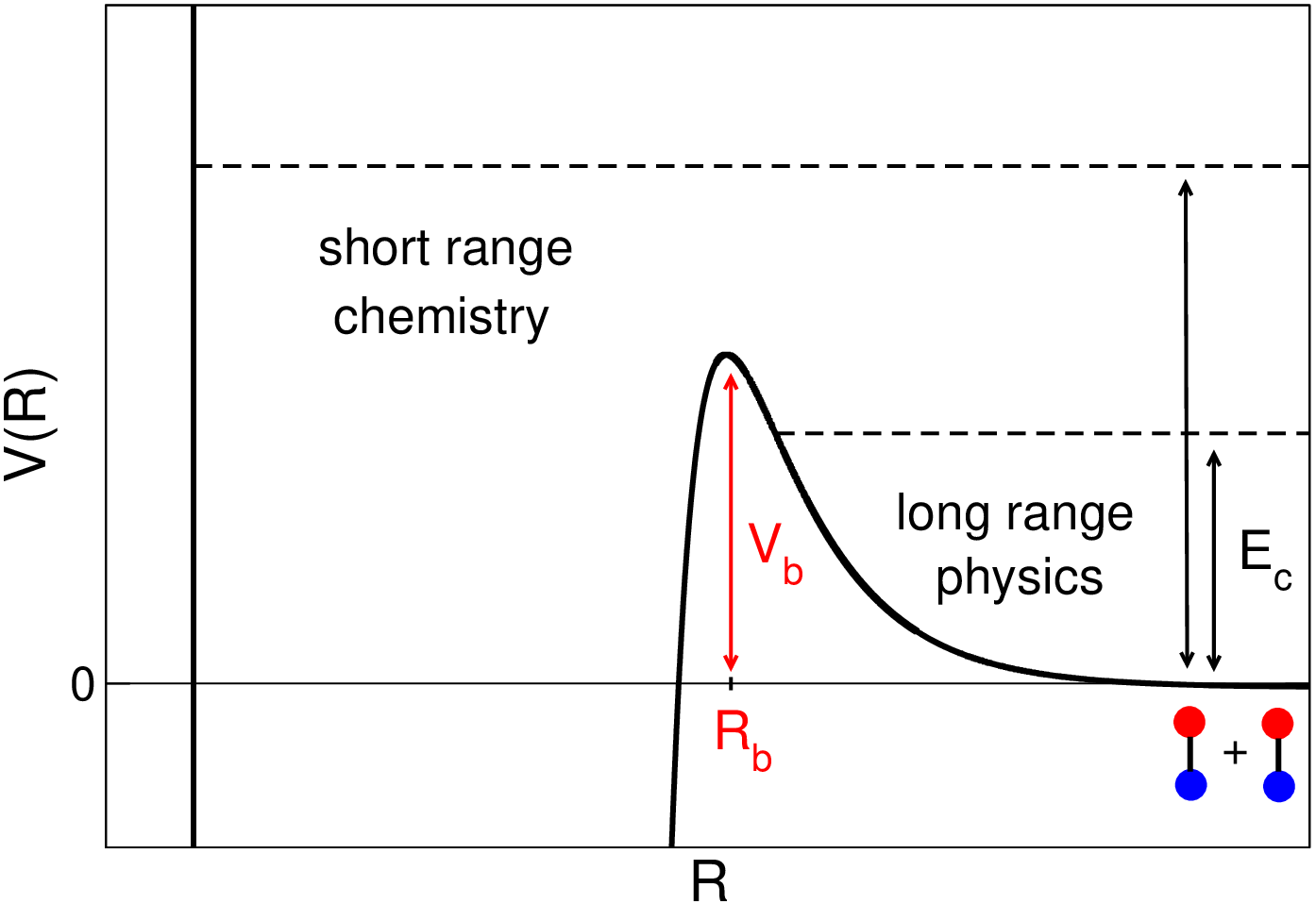}
\caption{
(Color online)
Effective potential barrier $V(R)$
as a function of the intermolecular separation $R$.
$V_b$ and $R_b$ denote the height and the position of the centrifugal barrier.
\label{spag-FIG}
}
\end{center}
\end{figure}

We consider the case of
two identical ultracold fermionic
polar molecules, 
as has been achieved very recently
for KRb dimers~\cite{Ni08,Ospelkaus09} in their ro-vibronic
($^1\Sigma,v=0,N=0$) ground state.
Under these circumstances, because of Fermi exchange symmetry, the
relative orbital angular momentum quantum number $L$ between the two
molecules must take odd values $L = 1,3,5,7 ...$.  
These molecules are polar molecules and can be controlled by an electric field $\cal E$.
In the usual basis set of
partial waves $|L M_L \rangle$,
the long-range behavior of two colliding polar molecules in a presence of an electric field
is governed by an interaction potential matrix whose elements are
%
\begin{multline}
\langle L M_L | V(R) | L' M_L' \rangle =  \\
\left\{ \frac{\hbar^2 L(L+1)}{2 \mu R^2} - \frac{C_6}{R^6} \right\} \delta_{L,L'} \, \delta_{M_L,M_L'}   \\
- \frac{{C_3}(L,L';M_L)}{R^3} \delta_{M_L,M_L'} 
\label{barrierpot-efield}
\end{multline}
%
where $R$ denotes the distance between the two molecules.
The diagonal elements represent effective potentials
for the colliding molecules and the non-diagonal elements represent
couplings between them.
The coefficient $C_6$ is the van der Waals coefficient,
assumed to be isotropic in the present treatment.
The $- {C_3} / R^3$ is the term corresponding 
to the electric dipole-dipole interaction 
expressed in the partial wave basis set
$\langle L M_L | V_{dd}(R,\theta,\varphi) | L' M_L' \rangle$ 
between two polarized molecules
in the electric field direction,
with 
$V_{dd}(R,\theta,\varphi)= d^2 (1 - 3 \cos^2{\theta}) / (4 \pi \varepsilon_0 \, R^3)$,
where $d=d({\cal E})$ is the induced electric dipole moment,
and $\theta,\varphi$ represent the relative orientation between the molecules.
In the basis set of partial waves, $C_3$ takes the form
\begin{multline}
{C_3}(L,L';M_L) 
=  \alpha(L,L';M_L) \, \frac{d^2}{4 \pi \varepsilon_0}   \\
= 2 \ (-1)^{M_L} \ \sqrt{2L+1} \ \sqrt{2L'+1}   \\
\ \left( \begin{array}{ccc} L & 2 & L' \\ 0 & 0 & 0 \end{array} \right)
\ \left( \begin{array}{ccc} L & 2 & L' \\ -M_L & 0 & M_L' \end{array} \right) \ \frac{d^2}{4 \pi \varepsilon_0} .
\label{C3coef}
\end{multline}
The large bracket symbols denote the usual 3$-j$ coefficients.
The coefficient $\alpha$ is introduced to simplify further notations.
The combination between repulsive and attractive terms
in the effective potentials (diagonal terms) of Eq.~\eqref{barrierpot-efield}
generate a potential barrier of height $V_b$
which is plotted schematically in Fig.~\ref{spag-FIG}.
The height of this barrier plays a crucial role as it can prevent
the molecule from accessing the short range region where reactive chemistry occurs.

The quantum threshold (QT) model consists of two conditions. 
First, for
$E_c < V_b$, we use the Bethe--Wigner threshold laws~\cite{Bethe35,Wigner48} for
ultracold scattering.
Second, we use the
classical capture model (Langevin model)~\cite{Langevin05}
to estimate the probability of quenching
for $E_c \ge  V_b$.
A classical capture model is indiferent to  collision energies
$E_c < V_b$ since the barrier prevents the molecules from coming close together.
In real-life quantum scattering, collisions
do occur at these energies due to quantum tunneling, and they are the ones relevant to
ultracold collisions.  Moreover, collisions in this energy regime
are dictated by the the Bethe--Wigner quantum threshold laws.
For quenching collisions, 
the threshold laws~\cite{Bethe35,Wigner48,Sadeghpour00} state that
$|T^{qu}_{L,M_L}|^2 \propto E_c^{L+\frac{1}{2}}$.
For $E_c \ge V_b$,
a classical capture model
will guarantee
to deliver the molecule pair to small
values of $R$, where chemical reactions are likely to occur
with unit probability (see Fig.~\ref{spag-FIG}).
Following this classical argument,
we will assume that when $E_{c} \ge  V_b $, the
quenching probability reaches unitarity $|T^{qu}|^2=1$.
Using this assumption together with the quantum threshold laws,
the QT quenching tunneling probability below the barrier can
be written as
\begin{eqnarray}
|T^{qu}_{L,M_L}|^2 = \left( \frac{E_c}{V_b} \right)^{L+1/2} . 
\label{Tqusq}
\end{eqnarray}
Consequently, the quenching cross section
and rate coefficient are approximated by
\begin{eqnarray}
\sigma^{qu}_{L,M_L} 
 & = & \frac{\hbar^2  \pi}{2 \mu V^{L+\frac{1}{2}}_b} \ E_c^{L-\frac{1}{2}}  \times \Delta \nonumber \\
K^{qu}_{L,M_L}  &=& 
 \frac{\hbar^2  \pi}{\sqrt{2 \mu^3} V_b^{L+\frac{1}{2}}} \  < E_c^{L} >   \times \Delta
\label{BetheWigner-n6}
\end{eqnarray}
for $E_c < V_b$.
The QT model has the simple and intuitive advantage 
of showing how the cross sections and rate coefficients 
scale with the height of the entrance centrifugal barrier.
For $E_c \ge V_b$,
it is easy to find the corresponding expression
of the cross section in Eq.~\eqref{cross} by setting $|T^{qu}|^2=1$.
The cross section $\sigma^{qu}_{L,M_L}$ will reach the unitarity
limit at $E_c \ge V_b$.
It is also easy to find the corresponding expression of the rate coefficient
in Eq.~\eqref{rate}.
The QT model is general for any collision between two particles
provided that there is a barrier in the entrance collision channel
and that chemical reactions occur with near unit probability at short range.
The only information on short
range chemistry is that chemical reactions occur at
full and unit probability and the only information on long range physics
is provided by the height of the entrance barrier $V_b$.
The QT model describes the background scattering process,
it does not take into account scattering resonances.
Note that the model will not be appropriate in the present form
for barrierless ($s-$wave) collisions
since $V_b=0$.
For this particular type of collisions
that do not possess a centrifugal barrier, the QDT theory can be usefully
applied~\cite{Ospelkaus-chemistry-09,Julienne09,Idziaszek09}.
The present form of the QT model 
does not take into account
the anisotropy
of the intermolecular potential
at intermediate range
and/or the electronic and nuclear spin structure of 
the molecular complex
but remains suitable
as far as the entrance centrifugal barrier
takes place at long range.
The QT model will have to be modified
if longer range interactions
takes place.
For example, collisions between $N=0$ and $N=1$
polar molecules can have long range interactions
between hyperfine states
due to dipolar and hyperfine couplings~\cite{Ospelkaus09,Aldegunde08}.
However, for collisions between rotationless $N=0$ polar molecules,
the hyperfine couplings are weak and the QT model 
can be applied without further modifications.

\subsection{Rates in zero electric field}

\begin{figure} [h]
\begin{center}
\includegraphics*[width=8cm,keepaspectratio=true,angle=0]{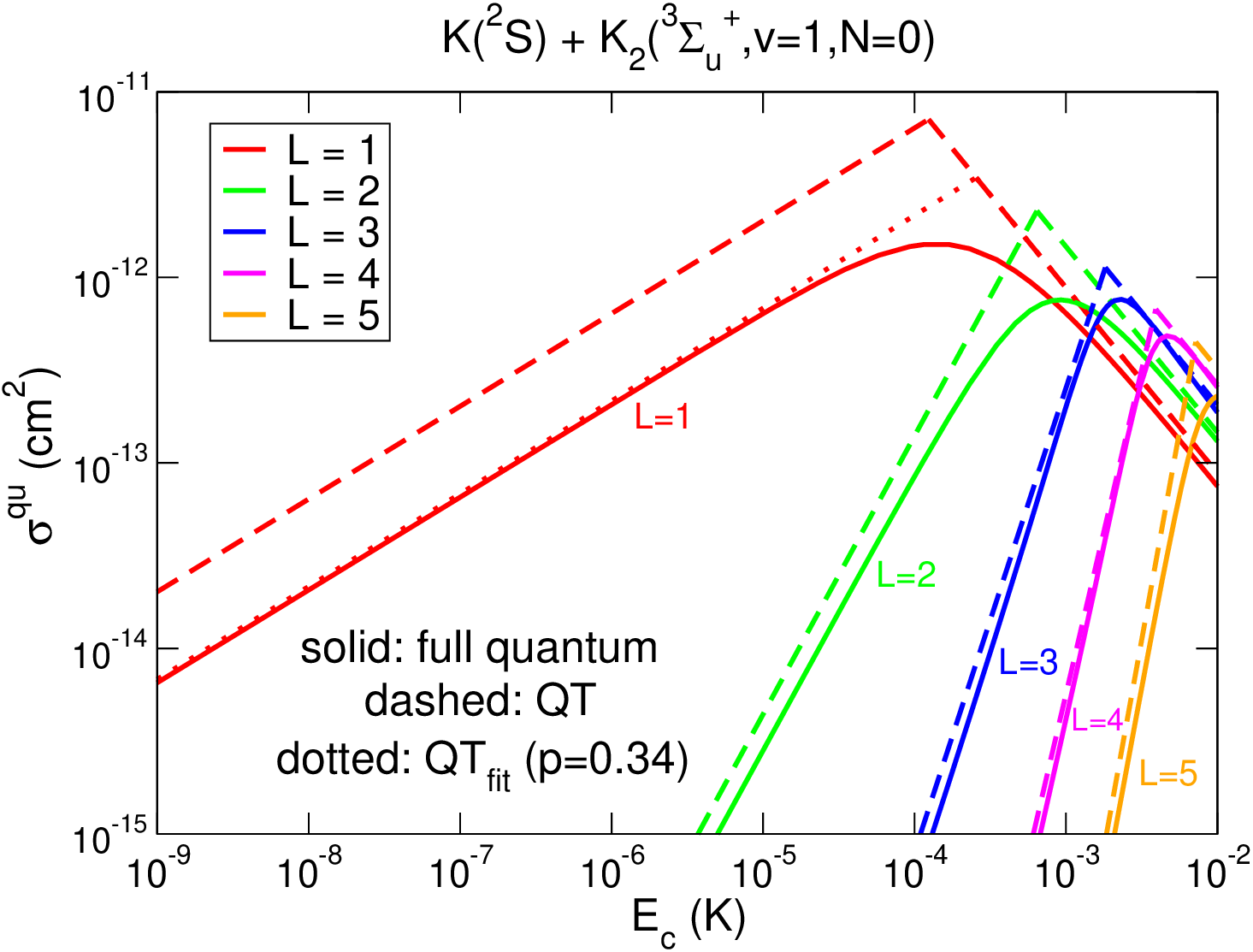}
\caption{
(Color online)
Quenching cross section
of $^{39}$K + $^{39}$K$_2$ as a function of the collision energy for the partial wave $L=1-5$:
(i) calculated with a full quantum calculation (solid lines), reproduced from Ref.~\cite{Quemener05}
(ii) using the QT model (dashed lines)
(iii) fitting the QT model (dotted line), using $p=0.34$ in Eq.~\eqref{Tqusqfit}.
\label{crossK3-FIG}
}
\end{center}
\end{figure}

\begin{figure} [h]
\begin{center}
\includegraphics*[width=8cm,keepaspectratio=true,angle=0]{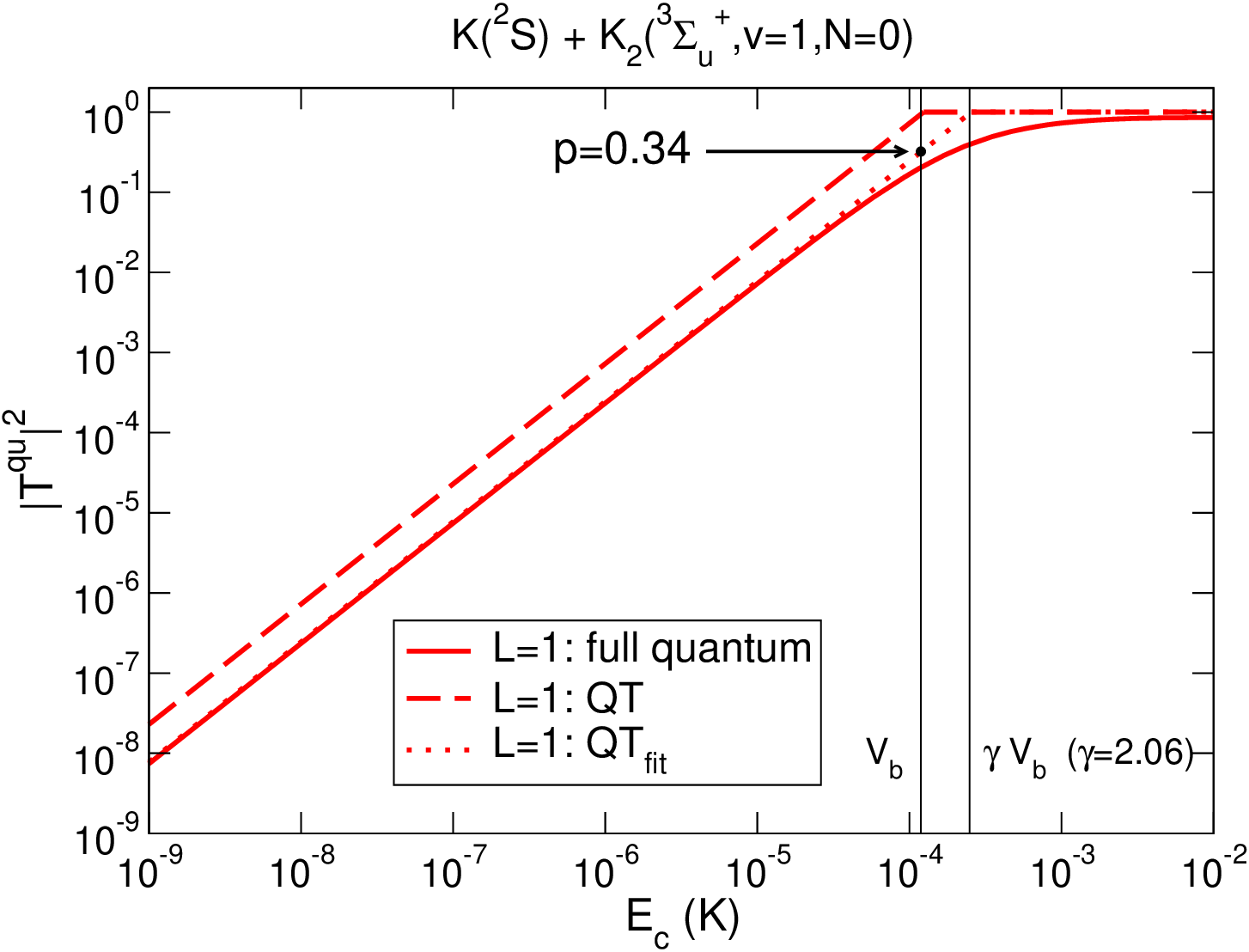}
\caption{
(Color online)
Quenching probability
of $^{39}$K + $^{39}$K$_2$ as a function of the collision energy for the partial wave $L=1$:
comparison between the full quantum calculation (solid line) and the QT model (dashed line).
The fitted QT model appears as a dotted line.
The height of the barrier $V_b$ and the corrected height $\gamma \times V_b$ ($\gamma=2.06$)
appear as vertical lines.
\label{probaK3-FIG}
}
\end{center}
\end{figure}

In the absence of an electric field in Eq.~\eqref{barrierpot-efield}, the long range potential
reduces to a diagonal term in the basis set of partial waves.
The position and height of the barrier are given by
\begin{eqnarray}
R_b &=& \left( \frac{6 \, \mu \, C_6}{\hbar^2 L(L+1)} \right)^{1/4} \nonumber \\
V_b &=& \left( \frac{ \left(\hbar^2 L (L+1)\right)^3}{54 \,  \mu^3 \,  C_6} \right)^{1/2} .
\label{Vn6}
\end{eqnarray}
We can insert Eq.~\eqref{Vn6} in Eq.~\eqref{BetheWigner-n6} to get analytical forms
of the quenching cross section or rate coefficient.
For two indistinguishable fermionic polar molecules at ultracold temperatures when $L=1$,
we get
\begin{eqnarray}
K^{qu}_{L=1,M_L} & = &
\frac{\pi}{8} \,
\left( \frac{3^{13} \, \mu^3 \, C_6^3}{\hbar^{10}} \right)^{1/4}
\ k_B T \times \Delta .
\label{ratequn6L1}
\end{eqnarray}
In Eq.~\eqref{BetheWigner-n6}, 
we used the fact that $<E_c> = 3 k_B T / 2$ in three dimensions.
Note that to get the overall contribution for a given $L$,
we have to multiply Eq.~\eqref{ratequn6L1} by the degeneracy factor $(2L+1)$ 
corresponding to all values of $M_L$.
We can get similar expressions for any partial wave $L$.

To test the validity of the model,
we compare in Fig.~\ref{crossK3-FIG} the quenching cross sections of 
$^{39}$K$(^2S)$ + $^{39}$K$_2(^3\Sigma_u^+,v=1,N=0)$ collisions
as a function of the collision energy
for the partial waves $L=1-5$:
(i) calculated in Ref.~\cite{Quemener05} with a full quantum
time-independent close-coupling calculation
based on hyperspherical democratic coordinates~\cite{Launay89}
and the full potential energy surface of K$_3$ (solid lines)
(ii) using the simple QT model (dashed lines) with a value of $C_6=9050$~a.u.
given in Ref.~\cite{Quemener05}
(1~a.u. = 1~$E_h \, a_0^6$
where $E_h$ is the Hartree energy and $a_0$ is the Bohr radius).
In this example, 
the QT model provides an upper limit to the cross sections. 
This is due to the fact that the quenching cross section does not reach
a maximum value at the height of the barrier $V_b$, but rather at somewhat higher energy, 
say $\gamma \times V_b$, with $\gamma > 1$ (see Ref.~\cite{Quemener05}).
For all partial waves, there is a worse agreement for collision energies in the vicinity of the height of the barrier
where the passage from the ultralow regime to the unitarity limit is smoother than for the QT model.
This smoother passage is visible in Fig.~\ref{probaK3-FIG} for the full quantum $L=1$ quenching probability
compared to the QT model, which has a sharp corner in the vicinity of $V_b$.

To account for more flexibility in the QT model, 
one can replace $V_b$ in Eq.~\eqref{Tqusq}
by $\gamma \times V_b$ ($\gamma > 1$), and use the coefficient $\gamma$
as a fitting parameter to reproduce either full quantum calculations
or experimental observed data. 
Alternatively, we can correct 
the QT quenching tunneling probability
with an overall factor $p$,
\begin{eqnarray}
{|T^{qu}_{L,M_L}|}^2_{\text{fit}} & = & p \times \left( \frac{E_c}{V_b} \right)^{L+1/2} 
\label{Tqusqfit}
\end{eqnarray}
with $p=\gamma^{-(L+1/2)}$.
$p<1$ can be interpreted
as the
quenching probability
reached at the height of the barrier $V_b$ in the QT model,
rather than the rough full unit probability ($p=1$).
As an example, we find that $\gamma \approx 2.06$
reproduces the quantum $L=1$ partial wave cross section
for $^{39}$K + $^{39}$K$_2$ (dotted line in Fig.~\ref{crossK3-FIG} and Fig.~\ref{probaK3-FIG}).
This yields a maximum quenching probability
of $p = \gamma^{-3/2} \approx 0.34$ instead of 1.
In other words, the QT model is only a factor 
of $p^{-1} \approx 2.96$ higher than the full quantum calculation
for $^{39}$K + $^{39}$K$_2$ collisions at ultralow energies.

Given the fact that
full quantum calculations
are computationally 
demanding~\cite{Soldan02,Quemener04,Quemener05,Cvitas05a,Cvitas05b,Lara06,Quemener07}
and impossible at the present time 
for alkali molecule-molecule collisions,
the accuracy of the QT model
is satisfactory
and can be a quick and powerful alternative
way to estimate orders of magnitude for
the scattering observables.
Besides, agreement between the QT model with experimental data 
or full quantum calculations is expected to be satisfactory for collisions
involving alkali species, because 
it is likely that short range quenching couplings
will dominate and lead to high quenching probability
in the region where the two particles are close together~\cite{QuemenerCHAPTER}.
Very recently, Eq.~\eqref{ratequn6L1} of the QT model
has been applied
for the evaluation of ultracold chemical 
quenching rate of collisions of two $^{40}$K$^{87}$Rb
molecules in the same internal quantum state, and provided
good agreement with the experimental data~\cite{Ospelkaus-chemistry-09}.

\subsection{Rates in non-zero electric field}

\subsubsection{Numerical expressions}

\begin{figure} [h]
\begin{center}
\includegraphics*[width=8cm,keepaspectratio=true,angle=0]{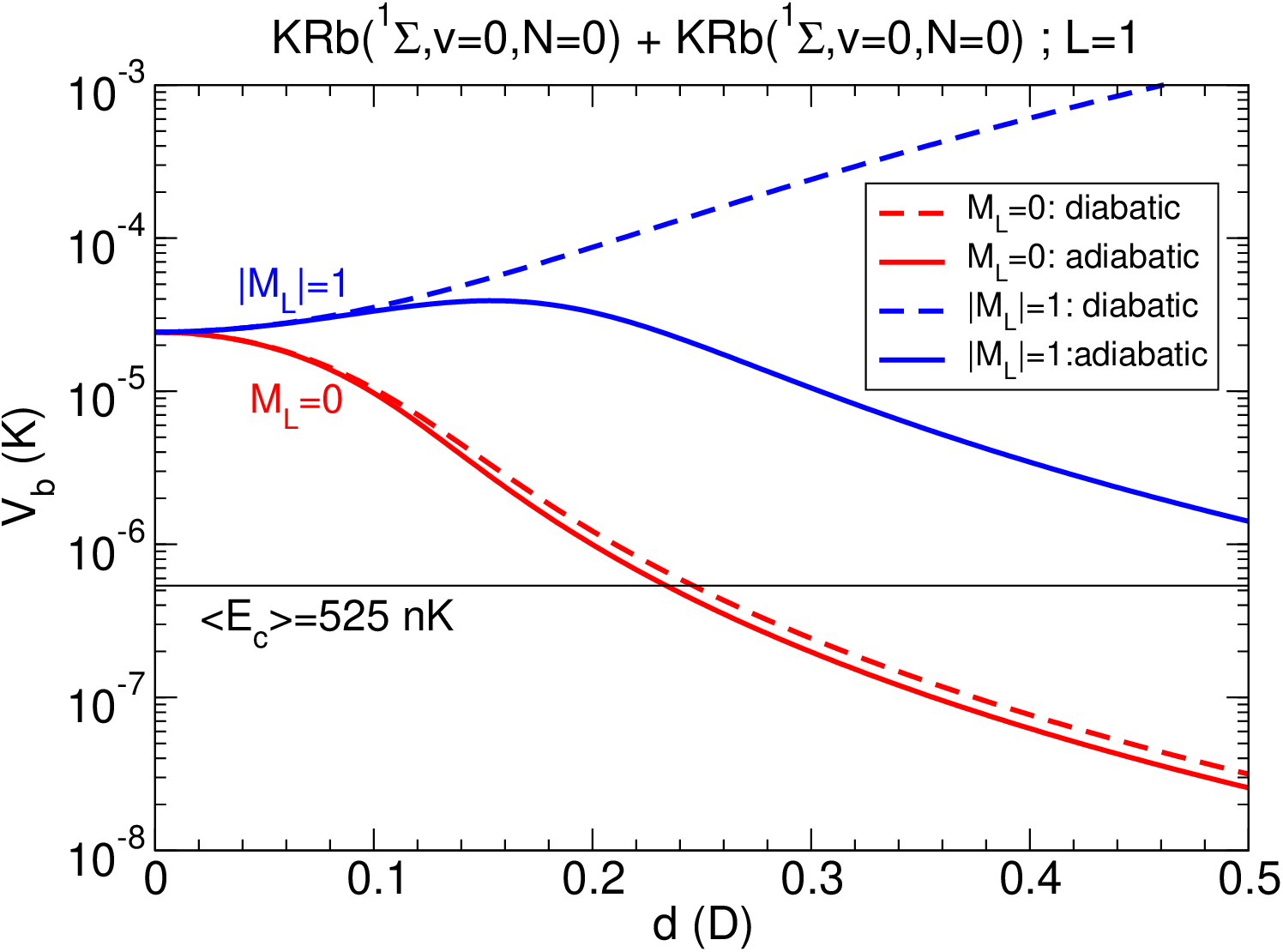}
\caption{
(Color online)
Diabatic (dashed lines)
and adiabatic (solid lines)
barrier heights $V_b^{d,a}$ as a function of the induced dipole moment $d$
for the partial waves $L=1, M_L=0$ (red curves) and $L=1, |M_L|=1$ (blue curves).
\label{barriers-FIG}
}
\end{center}
\end{figure}

\begin{figure} [h]
\begin{center}
\includegraphics*[width=8cm,keepaspectratio=true,angle=0]{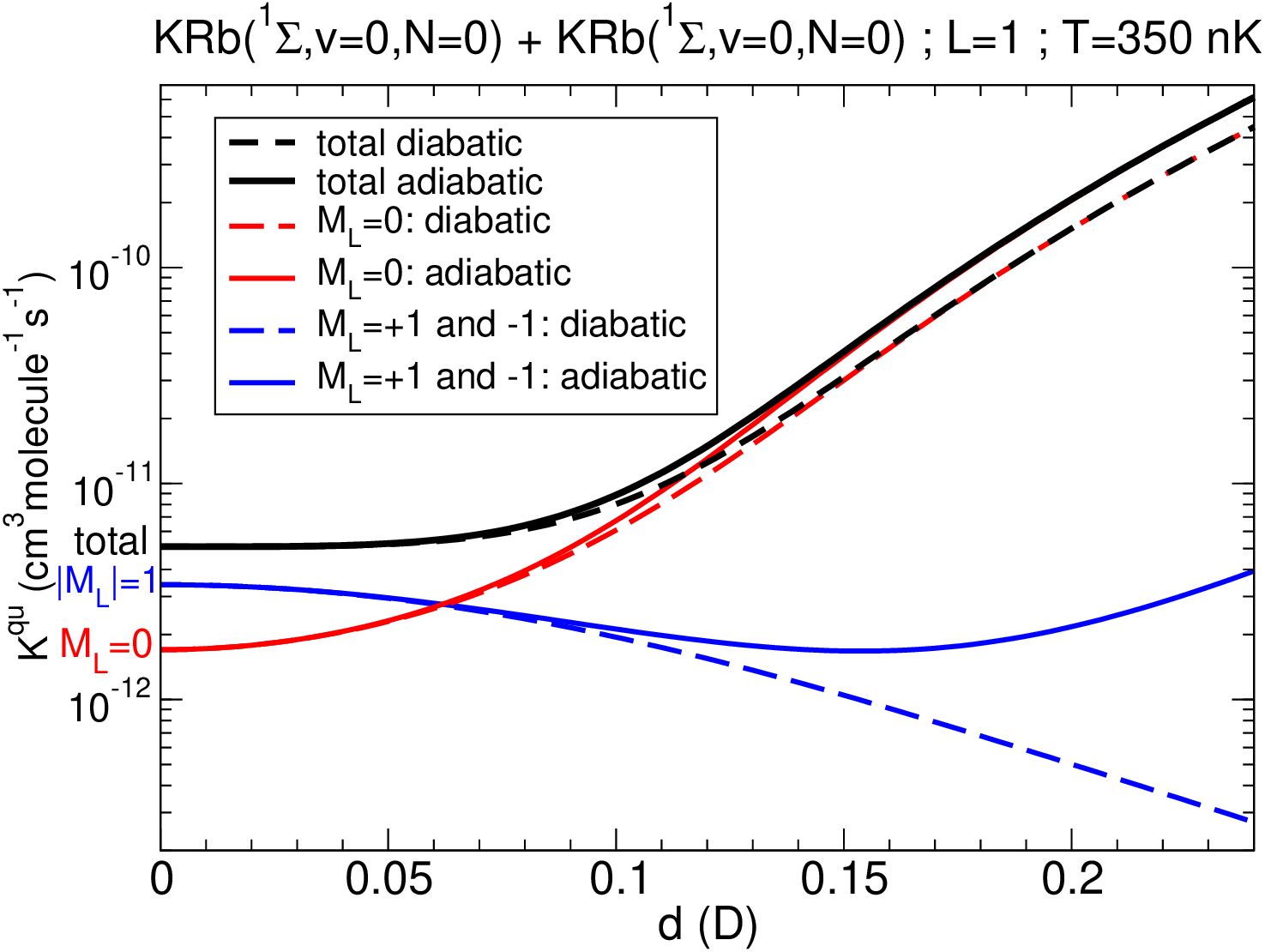}
\caption{
(Color online)
Quenching rate coefficients 
of two indistinguishable
fermionic polar $^{40}$K$^{87}$Rb  molecules
as a function of the induced electric dipole moment
for $L=1$ and for a temperature of $T=350$~nK (black curves).
The rates have been calculated using the barrier heights of Fig.~\ref{barriers-FIG}.
The red lines represent the $L=1, M_L=0$ partial wave contribution.
The blue lines represent the sum of $L=1, M_L=1$ and $L=1,M_L=-1$
partial wave contributions.
The dashed lines represent the rates calculated with the diabatic barriers
while the solid lines with the adiabatic barriers (see text for detail).
The total, $M_L=0$ and $|M_L|=1$ curves have been indicated in the left hand side.
\label{rate-num-FIG}
}
\end{center}
\end{figure}

In the presence of an electric field in Eq.~\eqref{barrierpot-efield}, the long-range interaction potential
matrix is no more diagonal and couplings between different values of $L$ occur. $M_L$ is still a good quantum number.
A first approximation (diabatic approximation)
consists of neglecting these couplings and using only the diagonal elements
of the diabatic matrix directly.
Then one can find numerically for which $R$ the centrifugal barriers
are maximum and evaluate the height of the diabatic barriers $V_b^d$.
This is repeated for all values of the induced dipole moment $d$.
A second approximation (adiabatic approximation)
is to diagonalize this matrix (including the non-diagonal coupling terms)
for each $R$ 
and again find the maximum of the centrifugal barriers
to get the height of the adiabatic barriers $V_b^a$.
As an example, we compute these barrier heights for $^{40}$K$^{87}$Rb$-$$^{40}$K$^{87}$Rb collisions, using
a value of $C_6 = 16130$~a.u.~\cite{Kotochigova09}.
We plot in Fig.~\ref{barriers-FIG} the heights of 
the diabatic (dashed lines) and adiabatic (solid lines) barriers
for the quantum numbers $M_L=0$ (red curves) and $|M_L|=1$ (blue curves).
The adiabatic barriers have been calculated using five partial waves $L=1-9$
in Eq.~\eqref{barrierpot-efield}.
The effect of the couplings can be clearly seen in this figure
by comparing diabatic and adiabatic barriers.
Especially for the $|M_L|=1$ case for $d \approx 0.16$~D
(1~D = 1~Debye = $3.336 \, 10^{-30}$~C.m),
couplings with higher partial waves
make the adiabatic barrier decrease
as the dipole increases while the diabatic barrier continues to increase.

Using these heights of the barriers, we use Eq.~\eqref{BetheWigner-n6}
to plot in Fig.~\ref{rate-num-FIG}
the total quenching rate coefficients (black curves) as a function of $d$
for two indistinguishable fermionic polar $^{40}$K$^{87}$Rb molecules in the same quantum state
for $L=1$ and at a typical experimental temperature of $T=350$~nK~\cite{Ospelkaus-chemistry-09}.
For $T=350$~nK, the mean collision energy $<E_c> = 3 k_B T /2 = 525$~nK,
and the maximum 
dipole moment for which
$V_b < 525$~nK (that is for which Eq.~\eqref{BetheWigner-n6} does not apply anymore) is around
$d \approx 0.24$~D (see Fig.~\ref{barriers-FIG}).
The dashed curves correspond to rates calculated with the diabatic approximation
while the solid curves correspond to rates calculated with the adiabatic approximation.
The $M_L=0$ contribution is plotted in red and the contribution of $M_L=+1$ and $M_L=-1$
is plotted in blue.
The rates highly reflect the behavior of the centrifugal barriers in the entrance collision channel.
When the barrier increases with the dipole,
it prevents the molecules from getting close together
and 
the quenching rates decreases.
When the barrier decreases, the tunneling probability
is increased allowing the molecules
to get close together,
and the quenching rates increases.

\subsubsection{Analytical expressions}

\begin{figure} [t]
\begin{center}
\includegraphics*[width=8cm,keepaspectratio=true,angle=0]{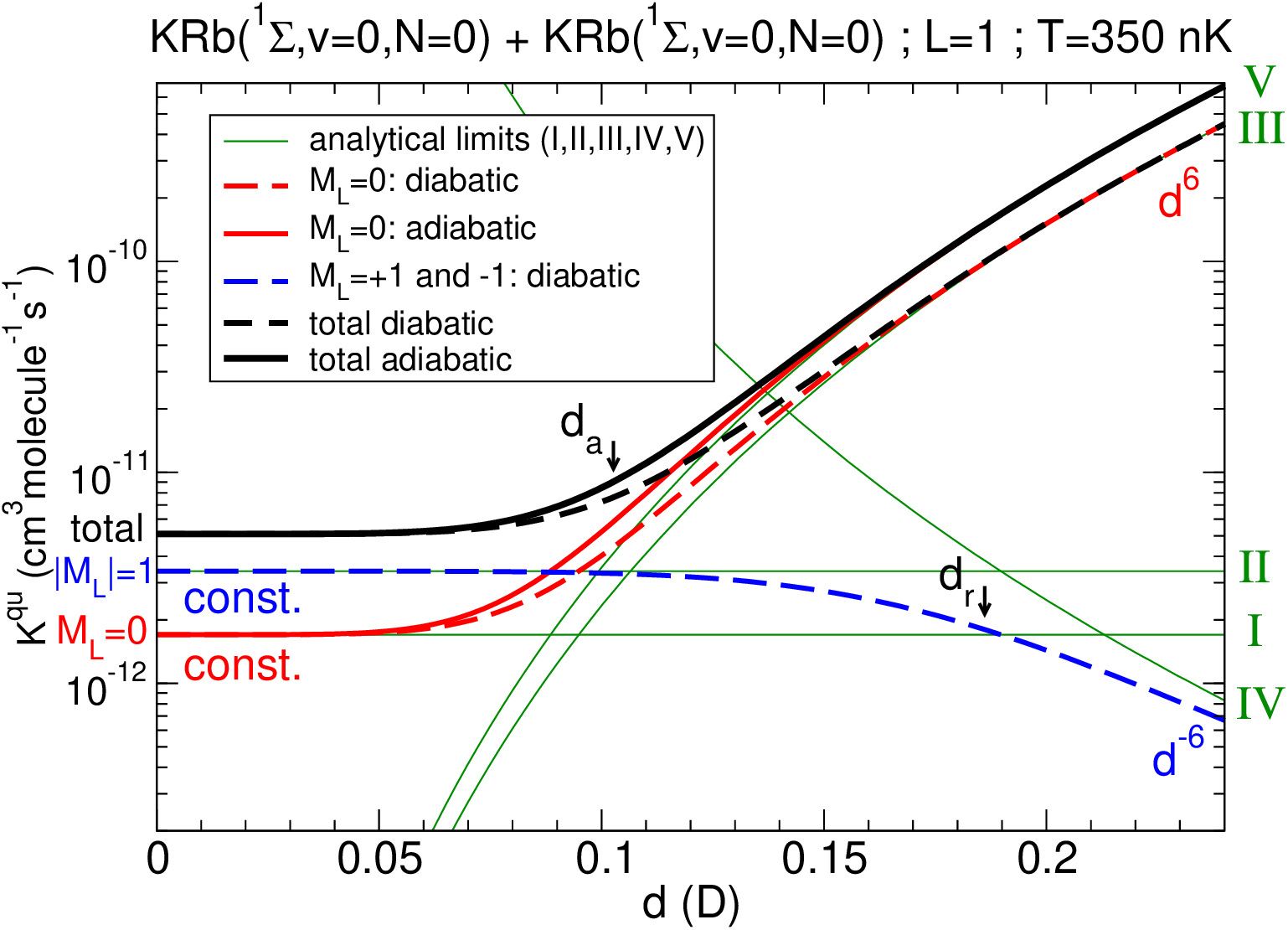}
\caption{
(Color online)
Same as Fig.~\ref{rate-num-FIG}
but we use the analytical expressions for the rates (see text for detail).
The total, $M_L=0$ and $|M_L|=1$ curves have been indicated in the left hand side.
The individual analytical curves have been indicated in the right hand side by roman numbers. 
\label{rate-anal-FIG}
}
\end{center}
\end{figure}

In order to have an intuitive
sense of how the chemical quenching rate scales
with the induced dipole moment (and the electric field),
we evaluate analytical expressions of the barriers and the rates as it has been done in
the previous section for a zero electric field.
The analytical expression of the height of the diabatic barrier $V_b^d$
is complicated by 
the occurrence of two distinct long-range potentials 
in the diagonal matrix term
of Eq.~\eqref{barrierpot-efield}.  We circumvent
this difficulty by looking in the two limits where one dominates
over the other.  For small electric fields, we use the zero electric field
limit discussed in the preceding section by setting $C_3 = 0$.
For larger electric fields we ignore the
$C_6$ coefficient in Eq.~\eqref{barrierpot-efield}
if the electric dipole-dipole interaction is attractive (positive $C_3$).
We ignore the centrifugal term in Eq.~\eqref{barrierpot-efield}
if the electric dipole-dipole interaction is repulsive (negative $C_3$).  
These two cases are discussed below.
In between, to accommodate the transition between the low-field and
high-field limit, we will simply add the rate coefficients derived in
the two limiting cases.

For positive $C_3$ coefficients in Eq.~\eqref{C3coef}, $-C_3/R^3$ is attractive
in Eq.~\eqref{barrierpot-efield}.
For
$L=1$ partial waves for example, this occurs when $M_L=0$, which favors an
attractive orientation of dipoles.
We consider 
\begin{eqnarray}
 \left|\frac{C_3}{R^3}\right| \gg \left|\frac{C_6}{R^6}\right|
\label{c3ggc6}
\end{eqnarray}
in Eq.~\eqref{barrierpot-efield}.
In this case, the position and height of the barrier are  given by
\begin{eqnarray}
R_b &=&  \frac{3 \, \mu \, C_3}{\hbar^2 L(L+1)}  \nonumber \\ 
V_b &=& \frac{\left(\hbar^2 L(L+1)\right)^{3}}{54 \, \mu^3 \, C_3^2}  \propto d^{-4} . 
\label{Vn3}
\end{eqnarray}
The position of the barrier in Eq.~\eqref{Vn3}
has to be in the region where Eq.~\eqref{c3ggc6} is satisfied.
This happens for suitably large dipole moments,   $d > d_a$ where
\begin{eqnarray}
d_a = \left( \frac{ \left( \hbar^2 L(L+1)\right)^3 \, C_6 \, (4 \pi \varepsilon_0)^{4} }{27 \, \mu^3 \, \alpha^4}\right)^{1/8} .
\end{eqnarray}
The subscript {\it a} stands for the attractive interaction.
For two indistinguishable fermionic polar $^{40}$K$^{87}$Rb molecules, 
and for $L=1$ and $M_L=0$, $\alpha(1,1;0)=4/5$ and
we get $d_a = 0.103$~D.
The threshold laws for quenching collisions in an electric field
are the same as in the zero-field limit.
Consequently, the quenching cross sections and rate coefficients
behaves as in Eq.~\eqref{BetheWigner-n6}
except that $V_b$ is given now by Eq.~\eqref{Vn3} and varies with $d$. 
We can insert Eq.~\eqref{Vn3} 
in Eq.~\eqref{BetheWigner-n6}
to get the corresponding analytical expressions.
For a partial wave $L>0$, 
the quenching rate  scales as $d^{4(L+\frac{1}{2})}$.
For two indistinguishable fermionic polar molecules at ultracold temperatures when $L=1$ and $M_L=0$,
we get
%
\begin{multline}
K^{qu}_{L=1,M_L=0}  =  \\ 
\frac{3 \pi}{8} \,
\left( \frac{6^9 \, \mu^6} {5^6 \, \hbar^{14}} \right)^{1/2}
\ \left(\frac{d^{2}}{4 \pi \varepsilon_0}\right)^{3}
\ k_B T \times \Delta .
\label{ratequn3L1}
\end{multline}
%
Thus the  $L=1, M_L=0$  quenching rate increases as $d^{6}$.  This is a more
rapid dependence on dipole moment than for purely long-range
dipolar relaxation in dipolar gases~\cite{Hensler03}.

For negative $C_3$ coefficients in Eq.~\eqref{C3coef}, $-C_3/R^3$ is repulsive
in Eq.~\eqref{barrierpot-efield}.  
For
$L=1$ partial waves for example, this occurs when $M_L=\pm 1$, which favors a
repulsive orientation of dipoles.
We consider 
\begin{eqnarray}
\left|\frac{C_3}{R^3}\right| \gg \left|\frac{\hbar^2 L(L+1)}{2 \mu R^2}\right| 
\label{c2ggc3}
\end{eqnarray}
in Eq.~\eqref{barrierpot-efield}.
The long-range potential again experiences a barrier, but now it is 
generated by the balance between the repulsive dipole potential at large $R$,
and the attractive van der Waals potential at somewhat smaller $R$.
In this case, the position and height of this barrier are given by
\begin{eqnarray}
R_b &=&  \left( \frac{2 \, C_6}{|C_3|} \right)^{1/3} \nonumber \\
 V_b &=& \frac{|C_3|^2} {4 \, C_6} \propto d^4  .
\label{Vn3rep}
\end{eqnarray}
For this approximation to hold, the position of the barrier in Eq.~\eqref{Vn3rep} 
has to be in the region where Eq.~\eqref{c2ggc3} is satisfied.
This requires that $d > d_r$ where
\begin{eqnarray}
d_r = \left( \frac{ \left(\hbar^2 L(L+1)\right)^3 \, C_6 \, (4 \pi \varepsilon_0)^{4} } {4 \, \mu^3 \, \alpha^4} \right)^{1/8} .
\end{eqnarray}
The subscript {\it r} stands for the repulsive interaction.
For two indistinguishable fermionic polar $^{40}$K$^{87}$Rb molecules, and
for $L=1$ and $M_L=1$ or $M_L=-1$,
$\alpha(1,1;\pm1)=2/5$ and
we get $d_r = 0.186$~D.
We can replace Eq.~\eqref{Vn3rep}
in Eq.~\eqref{BetheWigner-n6}
to get the corresponding analytical expressions.
For a partial wave $L>0$,
the quenching processes scale as $d^{-4(L+\frac{1}{2})}$.
For two indistinguishable fermionic polar molecules at ultracold temperatures when $L=1$ and $|M_L|=1$,
we get
%
\begin{multline}
K^{qu}_{L=1,|M_L|=1}  =  \\
\frac{3 \pi}{8}
\left( \frac{50 \, \hbar^{\frac{4}{3}} \, C_6}{\mu} \right)^{3/2}
\ \left(\frac{d^{2}}{4 \pi \varepsilon_0}\right)^{-3}
\ k_B T \times \Delta.
\label{ratequn3L1rep}
\end{multline}
%
The $L=1, |M_L|=1$ quenching rate decreases as $d^{-6}$
as the electric field grows.

These analytical expressions use the diabatic barriers.
If we consider that at large $d$, the total rate is mostly given by the $M_L=0$
contribution (we neglect the $|M_L|=1$ contributions at large $d$), 
one can have an analytical expression using the adiabatic barrier.
If we take into account the couplings between $L=1, M_L=0$ and $L=3, M_L=0$,
we can diagonalize analytically the $2\times2$ matrix in Eq.~\eqref{barrierpot-efield}.
It can be shown that for each dipole moment $d$, 
the coupling with $L=3,M_L=0$ lower the diabatic barrier of $L=1,M_L=0$ by a factor of 0.76
at the position of the barrier,
to give rise to the adiabatic barrier.
Inserting this correction of the barrier in Eq.~\eqref{BetheWigner-n6}, this yields a correction
of $0.76^{-3/2} \approx 1.51$ for $K^{qu}_{L=1,M_L=0}$.
The difference between diabatic and adiabatic calculations
can be already seen in Fig.~\ref{barriers-FIG} 
and Fig.~\ref{rate-num-FIG}
for the numerical barriers
at large dipole moment.

In Fig.~\ref{rate-anal-FIG}
the black curve corresponds
to the total quenching rate coefficient as a function of $d$
for two indistinguishable fermionic polar $^{40}$K$^{87}$Rb molecules in the same quantum state
for $L=1$ and at a temperature of $T=350$~nK.
The analytical expressions I, II, III, IV, V (green thin lines)
correspond respectively to
Eq.~\eqref{ratequn6L1}, 2~$\times$~Eq.~\eqref{ratequn6L1}, Eq.~\eqref{ratequn3L1}, 2~$\times$~Eq.~\eqref{ratequn3L1rep}, 1.51~$\times$~Eq.~\eqref{ratequn3L1}.
The curves III and IV are for the diabatic barriers, while curve V is to account for the adiabatic barrier.
The red dashed line (I+III) represents the $L=1, M_L=0$ partial wave contribution for the diabatic barriers 
while the blue dashed line (II+IV) represents the sum of $L=1, M_L=1$ and $L=1,M_L=-1$
partial wave contributions for the diabatic barriers.
The analytical sum I+II+III+IV is represented as a black dashed line.
To account for the adiabatic barriers
we assume that the correction for the total rate
comes only from the $L=1, M_L=0$ partial wave, 
and we replace I+III by I+V (red solid line).
The analytical sum I+II+V+IV is represented as a black solid line.
Neglecting the $d^{-6}$ contribution at larger $d$,
the analytical $p-$wave quenching rate (taking into account
the adiabatic barriers) is given by the simple expression
\begin{multline}
K^{qu}_{L=1} = 
\frac{\pi}{8} \,  
\bigg\{
p_1 
\left( \frac{3^{17} \, \mu^3 \, C_6^3}{\hbar^{10}} \right)^{1/4} \\
+
1.51 \, p_2 \left( \frac{2^9 \, 3^{11} \, \mu^6} {5^6 \, \hbar^{14}} \right)^{1/2}
\, \frac{d^{6}}{(4 \pi \varepsilon_0)^{3}}
\bigg\}  
\, k_B T \times \Delta.
\label{rateanal}
\end{multline}
$p_1$ ($p_2$) is the quenching probability reached at the height of the barrier in the QT model
for the zero (non-zero) electric field regime.
The QT model assumes that $p_1=p_2=1$ but become fitting parameters
($p_1,p_2 < 1$) when compared with full quantum calculations or experimental data.
The limiting value $d_a=0.103$~D ($d_r=0.186$~D), 
where the $d^6$ ($d^{-6}$) behavior begins, has also been indicated with an arrow.
It turns out that the total rates for $L=1$ 
calculated analytically (for both the use of diabatic and adiabatic barriers)
are very similar to the numerical ones of Fig.~\ref{rate-num-FIG} (10~\% difference
at most, around $d_a$).
However, the sub-components $L=1, M_L=0$ and $L=1, |M_L|=1$  
have different behaviors.
For example the numerical $L=1, M_L=0$ ($L=1, |M_L|=1$) component starts 
to increase (decrease) at earlier dipole moment (typically at 0.02~D)
than their analytical analogs (typically after 0.06~D).
The use of the simple analytical expressions (using the diabatic or adiabatic barriers)
can be useful to estimate the total rate coefficients, while
the numerical ones are prefered to estimate
the $L=1, M_L=0$ and $L=1, |M_L|=1$ individual rates.

\section{Prospects for collisions in two dimensions}

\begin{figure} [t]
\begin{center}
\includegraphics*[width=8cm,keepaspectratio=true,angle=0]{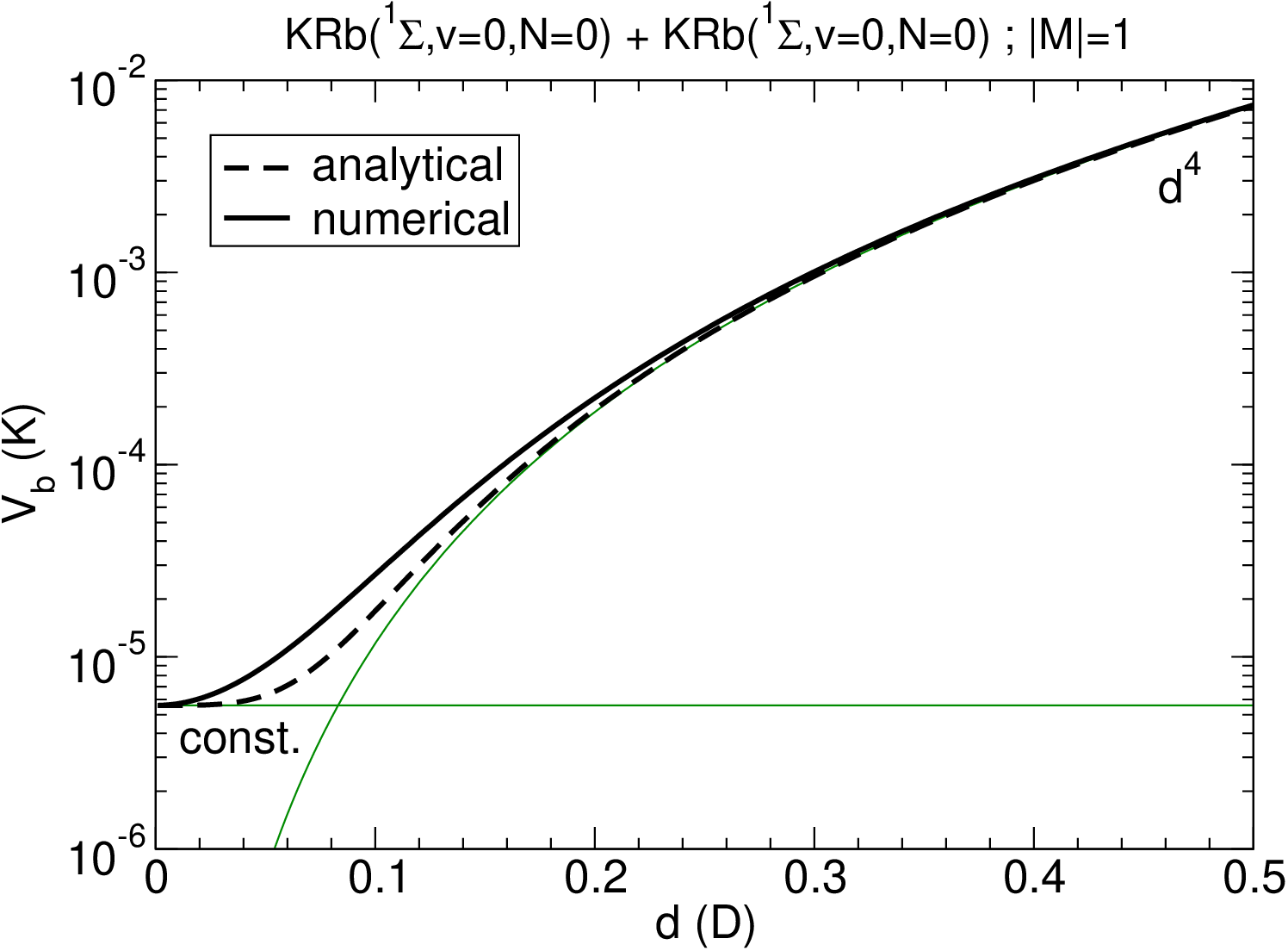}
\caption{
(Color online)
Barrier heights as a function of the induced dipole moment $d$
for the partial waves $|M|=1$ in two dimensions.
The green thin curves represent the analytical Eq.~\eqref{barrier2D-const} (constant)
and Eq.~\eqref{barrier2D-d4} ($d^4$).
The dashed black curve is the sum of them.
The solid black curve is the height of the barrier in Eq.~\eqref{barrierpot-efield-2D}.
\label{barriers2D-FIG}
}
\end{center}
\end{figure}

\begin{figure} [h]
\begin{center}
\includegraphics*[width=8cm,keepaspectratio=true,angle=0]{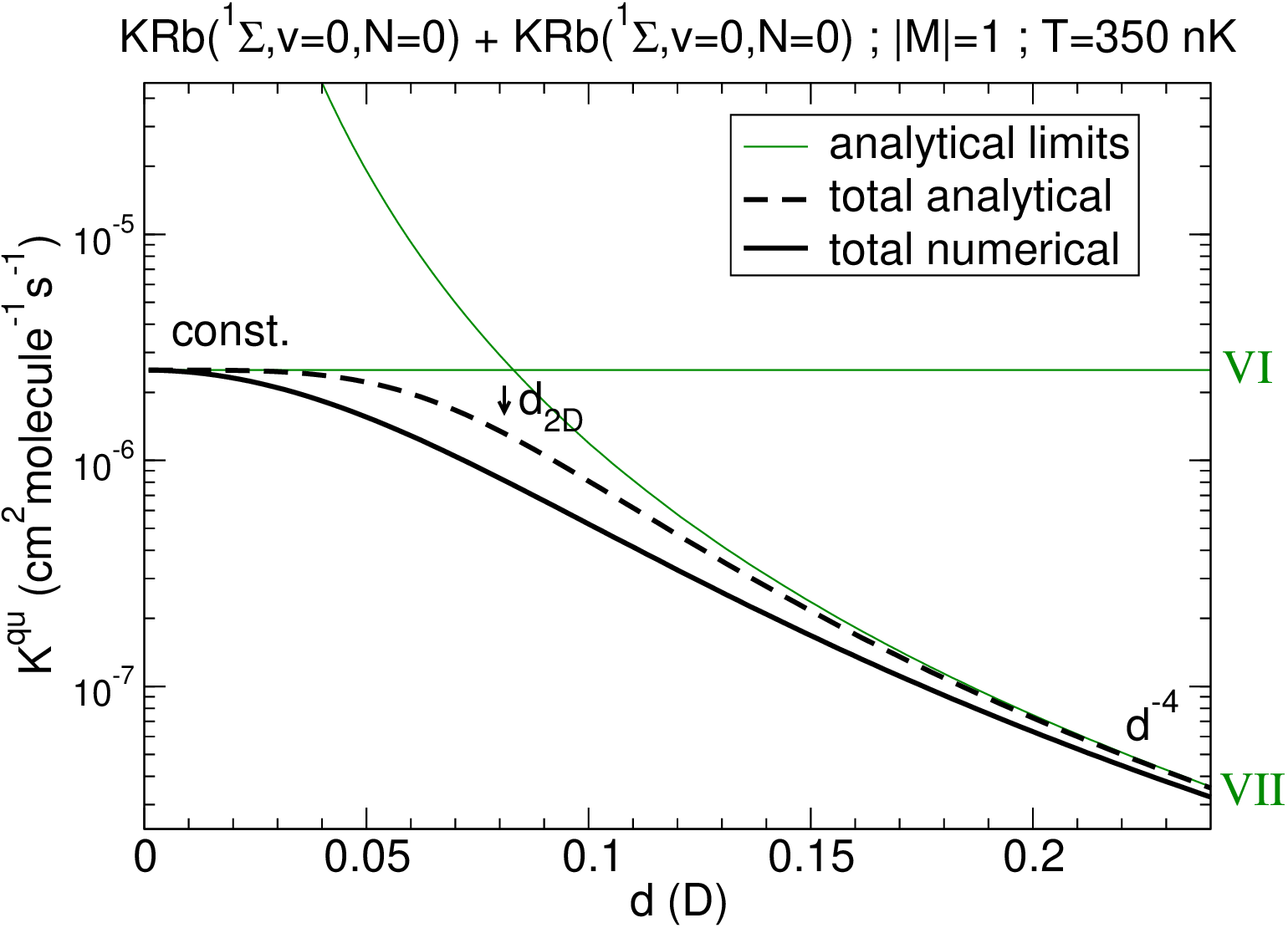}
\caption{
(Color online)
Two dimensional quenching rate coefficient (black curves)
of two indistinguishable fermionic polar $^{40}$K$^{87}$Rb molecules
as a function of the induced electric dipole moment
for the $M=1$ and $M=-1$ components at a temperature of $T=350$~nK.
The dashed lines represent the rate using analytical expressions
while the solid line represents the rate using the numerical expression (see text for detail). 
The individual analytical curves have been indicated in the right hand side by roman numbers. 
\label{2D-FIG}
}
\end{center}
\end{figure}

In three dimensional collisions, 
the quenching loss is largely due to incident partial waves with
angular momentum projection $M_L=0$, emphasizing head-to-tail orientations of pairs of dipoles.  
These are the kind of collisions 
that are largely suppressed in traps with a pancake geometry, with the dipole 
polarization axis orthogonal to the plane of the pancake~\cite{Micheli07}.  
If these collisions can be removed, then it is likely that increasing the
electric field will suppress quenching collisions, making evaporative cooling
possible.
If we assume an ideal pancake trap that confines
the particles to move strictly on a plane,
one can apply the present model to estimate 
the behavior of the quenching processes.
We assume that the molecules are polarized along the electric field axis, 
perpendicular to the two dimensional plane.
In this case,
the long range potential is given by 
\begin{eqnarray}
V(\rho) =  \frac{\hbar^2 (|M|^2 - 1/4)}{2 \mu \rho^2} - \frac{C_6}{\rho^6}  
+ \frac{d^2}{4 \pi \varepsilon_0 \, \rho^3} 
\label{barrierpot-efield-2D}
\end{eqnarray}
where $\rho$ stands for the distance between 2 particles in a two dimensional plane,
$M$ stands for the angular momentum projection on the electric field axis.
The last term comes from the repulsive dipole-dipole interaction 
when the dipoles are pointing along the electric field
and approach each other side by side.
The height of this barrier has been plotted as 
a function of $d$ in Fig.~\ref{barriers2D-FIG} (black solid line). 
At ultralow energy and large molecular separation, 
the Bethe--Wigner laws for quenching processes depend only on the long-range repulsive centrifugal term $1/R^2$~\cite{Sadeghpour00,Rau84}.
The repulsive centrifugal terms are different in Eq.~\eqref{barrierpot-efield}  and Eq.~\eqref{barrierpot-efield-2D}.
As the repulsive centrifugal term in Eq.~\eqref{barrierpot-efield} leads to the threshold laws in Eq.~\eqref{Tqusq},
the replacement $L(L+1) \to |M|^2-1/4$ (that is $L \to |M|-1/2$) in Eq.~\eqref{Tqusq} leads to
\begin{eqnarray}
|T^{qu}_{M}|^2 = \left( \frac{E_c}{V_b} \right)^{|M|} 
\label{Tqusq-2D}
\end{eqnarray}
where $V_b$ denotes the height of the centrifugal barrier in two dimensions.
This result requires that the centrifugal potential is repulsive,
i.e., that $|M|>0$. For $M=0$ the threshold law exhibits instead a
logarithmic divergence~\cite{Sadeghpour00}.

In two dimensions, quenching cross sections and rate coefficients 
have respectivelly units of length and length squared per unit of time, and are given by~\cite{Naidon06}
\begin{eqnarray}
\sigma^{qu}_{M} &=&  \frac{\hbar}{\sqrt{2 \mu E_c}} \ |T^{qu}_{M}|^2  \times \Delta  \nonumber \\ 
K^{qu}_{M}  
&=& \frac{\hbar}{\mu} \ |T^{qu}_{M}|^2 \times \Delta.
\label{cross-2D}
\end{eqnarray}
Within this model, it follows that the quenching cross section and rate coefficient for $|M| > 0$ are given by
\begin{eqnarray}
\sigma^{qu}_{M} 
 & = & \frac{\hbar}{\sqrt{2 \mu} \, V_{b}^{|M|}} \ E_c^{|M|-1/2}  \times \Delta \nonumber \\
K^{qu}_{M}  &=& 
 \frac{\hbar}{\mu \, V_{b}^{|M|}} \  < E_c^{|M|} >   \times \Delta .
\label{crossrate2D}
\end{eqnarray}
The energy dependence is in agreement with the one found
in Ref.~\cite{Li09}.
In Eq.~\eqref{crossrate2D},
\begin{eqnarray}
V_{b} = \left(\frac{(\hbar^2 (|M|^2 - 1/4))^3}{54 \, \mu^3 \, C_6}\right)^{1/2} 
\label{barrier2D-const}
\end{eqnarray}
for the zero-electric field regime
and
\begin{eqnarray}
V_{b} = \frac{1}{4 \, C_6} \left(\frac{d^2}{4 \pi \varepsilon_0}\right)^2 
\label{barrier2D-d4}
\end{eqnarray}
for the non-zero electric field regime.
The height of these barriers
has been reported in Fig.~\ref{barriers2D-FIG} (green thin lines).
These results imply that for $|M|=1$ 
the quenching processes
within this model
will be independent of the dipole moment in the zero electric field regime,
where
\begin{eqnarray}
K^{qu}_{|M|=1} & = &
\left( \frac{2^{7} \, \mu \, C_6 }{\hbar^4} \right)^{1/2}
\ k_B T \times \Delta
\label{rate2D-c6}
\end{eqnarray}
and
will scale as 
$d^{-4}$ 
in the non-zero electric field regime, where
\begin{eqnarray}
K^{qu}_{|M|=1} & = &
\frac{4 \, \hbar \, C_6}{\mu} 
\ \left(\frac{d^{2}}{4 \pi \varepsilon_0}\right)^{-2}
\ k_B T \times \Delta.
\label{rate2D-c3}
\end{eqnarray}
We use the fact that $<E_c> = k_B T$
in two dimensions.
The non-zero electric field regime is reached when $d > d_{2D}$ where
\begin{eqnarray}
d_{2D} = \left( \frac{ \left(\hbar^2 (|M|^2 - 1/4) \right)^3 \, C_6 \, (4 \pi \varepsilon_0)^{4}} {4 \, \mu^3} \right)^{1/8}
\end{eqnarray}
For two indistinguishable fermionic polar $^{40}$K$^{87}$Rb molecules, and
for $|M|=1$,
we get $d_{2D} =0.081$~D.

The behavior of the quenching rate (black lines) is shown in Fig.~\ref{2D-FIG}
for two indistinguishable fermionic polar $^{40}$K$^{87}$Rb molecules
as a function of the induced electric dipole moment
for $M=1$ and $M=-1$ components at a temperature of $T=350$~nK.
The dashed line represents the analytical rate 
which is the sum of the analytical expression VI corresponding to
$2 \times $~Eq.~\eqref{rate2D-c6}
and analytical expression VII
corresponding to 
$2 \times $~Eq.~\eqref{rate2D-c3}.
The solid line represents the rate using the general expression Eq.~\eqref{crossrate2D}
and the numerical height of the barrier calculated in Eq.~\eqref{barrierpot-efield-2D}.
The limiting value $d_{2D} = 0.081$~D, where the $d^{-4}$ behavior for the quenching rate begins, 
has also been indicated with an arrow.
The difference between the numerical calculation 
and the analytical expression 
reflects
the difference in the calculation of the height of the barrier,
already seen in Fig.~\ref{barriers2D-FIG}.
The numerical calculation
is more exact, while the other is analytical.
However, at large $d$, the numerical rate tends to the analytical $d^{-4}$ behavior.
The quenching rate decreases rapidly as the dipole moment increases
and this may be promising for efficient evaporative cooling of polar molecules
since the elastic rate
is expected to grow with increasing dipole moment~\cite{Ticknor09}.

\section{Conclusion}

We have proposed a simple model which combines quantum threshold laws and
a classical capture model to determine analytical expressions
of the chemical quenching cross section and rate coefficient
as a function of the collision energy or the temperature. 
We also provide an estimate as a function of 
the induced electric dipole moment $d$ in the presence of an electric field.
We found that
the quenching rates of two ultracold indistinguishable fermionic polar molecules
grows as the sixth power of $d$.
For weaker electric field, quenching processes
are independent of the induced electric dipole moment.
Prospects for two dimensional collisions have been discussed
using this model and we predict
that the quenching rate will decrease as the inverse of the fourth power
of $d$. This fact may be useful for efficient evaporative cooling of polar molecules.
This model provides a general and comprehensive picture
of ultracold collisions in electric fields.
Preliminary data suggest that this model gives good agreement 
with experimental chemical rates for three dimensional collisions
in an electric field~\cite{Ni-dipolar-09}.

\section{Acknowledgements}

We acknowledge the financial support 
of NIST, the NSF, and an AFOSR MURI grant.
We thank
K.-K. Ni, S. Ospelkaus,
D. Wang, M. H. G. de Miranda, B. Neyenhuis, P. S. Julienne,
J. Ye, and D. S. Jin for helpful discussions.

\end{document}